\documentstyle[preprint,aps,epsf]{revtex}
 
\newcommand {\be}{\begin{equation}}
\newcommand {\ee}{\end{equation}}
\newcommand {\bey}{\begin{eqnarray}}
\newcommand {\eey}{\end{eqnarray}}

\begin{document}
\draft

\title{CHRONOTOPIC LYAPUNOV ANALYSIS:\\
(II) TOWARDS A UNIFIED APPROACH}

\author{Stefano Lepri$^{1,2}$, Antonio Politi$^{2,3}$ and Alessandro
Torcini$^4$}

\address{
1 {\it Dipartimento di Fisica, Universit\'a di Bologna \\and 
Istituto Nazionale di Fisica Nucleare, I-40127 Bologna, Italy} \\
2 {\it Istituto Nazionale di Ottica I-50125 Firenze, Italy}\\
3 {\it Istituto Nazionale di Fisica della Materia\\
and Istituto Nazionale di Fisica Nucleare, 
I-50125 Firenze, Italy}\\
4 {\it Theoretische Physik, Bergische
Universit\"at-Gesamthochshule Wuppertal\\
D-42097 Wuppertal, Germany}}

\date{\today}
 
\maketitle
\begin{abstract}
>From the analyticity properties of the equation
governing infinitesimal perturbations, it is shown that all stability
properties of spatially extended 1D systems can be derived from a single
function that we call entropy potential since it gives directly the
Kolmogorov-Sinai entropy density. Such a function allows determining also
Lyapunov spectra in reference frames where time-like and space-like
axes point in general directions in the space-time plane. 
The existence of an entropy potential implies that the integrated
density of positive exponents is independent of the reference frame.
\end{abstract}
\vskip 1 true cm

\section{Introduction}

The first part of this work \cite{lyap1} (hereafter referred to as LPT) was
devoted to the definition and discussion of the properties of temporal (TLS) 
and spatial (SLS) Lyapunov spectra of 1D extended dynamical systems. 
In this second part we first show
how the two approaches are mutually related by proving, in some simple
cases, and conjecturing, in general, that all stability properties can be
derived from a single observable:  the {\it entropy potential}, which is
a function of two independent variables, the spatial and the temporal growth
rates $\mu$, $\lambda$, respectively. Legendre transforms represent the
right tool to achieve a complete description of linear stability properties 
in the space-time plane. In fact, we find that equivalent descriptions can be 
obtained by choosing any pair of independent variables in the set
$\{n_\mu,n_\lambda,\mu,\lambda\}$, where $n_\mu$ and $n_\lambda$ are the
integrated densities of spatial, resp. temporal, Lyapunov exponents. The
corresponding potentials are connected via suitable Legendre transformations
involving pairs of conjugated variables.

A further type of connection between spatial and temporal Lyapunov exponents
discussed in this paper is found in connection with the evolution of 
perturbations along generic ``world-lines'' in the space-time, i.e. along 
directions other than the space and time axes considered in LPT. The extension 
of the usual definition of Lyapunov exponents to this more general class of frames,
already discussed in \cite{bidime}, is rather appropriate for characterizing
patterns with some anisotropy. Here, we show that this seemingly more general 
class of {\it spatiotemporal} exponents can be derived from the knowledge of 
spatial and temporal Lyapunov spectra, which thus confirm to contain all the 
relevant information. 

For the sake of completeness, finally, we recall the last class of exponents 
introduced to describe convectively unstable states, {\it comoving} Lyapunov 
exponents \cite{como}, and their relationship again with SLS and TLS. In 
particular, we discuss the structure of the spectra in a simple case of a 
stationary random state.

Let us now briefly introduce the notations with reference to some specific 
models. Spatiotemporal chaos and instabilities in extended systems have been 
widely studied with the aid of models of reaction-diffusion processes, whose
general 1D form is of the type \cite{mannev,croho}
\be
\label{pde}
  \partial_t {\bf y} = {\bf F}({\bf y}) + {\bf D} \partial_x^2 {\bf y} \quad ,
\ee
with the state variable ${\bf y}(x,t)$ defined on the domain $[0,L]$
(periodic boundary conditions ${\bf y}(0,t)={\bf y}(L,t)$ are generally
assumed). The nonlinear function ${\bf F}$ accounts for the local 
reaction dynamics, while the diffusion matrix ${\bf D}$ represents 
the strength of the spatial coupling.
The introduction of coupled map lattices (CML) has been 
of great help for understanding the statistical properties of 
spatio-temporal chaos, especially  by means of numerical simulations.

In its standard form  \cite{kaneko,kapral} a CML dynamics reads
as
\be
\label{mappa}
  y^i_{n+1} = f\left((1-\varepsilon)y^i_n+ {\varepsilon \over 2}
      \left[  y^{i-1}_n+y^{i+1}_n \right]\right)\\ ,
\ee
where $i,n$ being the space, resp. time, indices labelling each variable 
$y^i_n$ of a lattice of length $L$ (with periodic boundary conditions 
$y_n^{i+L}=y^i_n$), and $\varepsilon$ gauges the diffusion strength. 
The function $f$, mapping a given interval $I$ of the real axis onto itself,
simulates a local nonlinear reaction process. 

A generalization of model (\ref{mappa}) has been proposed \cite{como} 
to mimic 1D open-flow systems, namely 
\be
\label{amappa}
  y^i_{n+1} = f\left((1-\varepsilon)y^i_n+\varepsilon 
      \left[ (1-\alpha) y^{i-1}_n+\alpha y^{i+1}_n \right]\right)\\.
\ee
The parameter $\alpha$ (bounded between 0 and 1) accounts for the 
possibility of an asymmetric coupling, corresponding to first order 
derivatives in the continuum limit. 

The present paper is organized as follows. In Section II
we introduce the entropy potential and derive its explicit expression
in some simple cases. Sec. III is devoted to Lyapunov analysis
in tilted reference frames, while Sec. IV deals with the relationships
between temporal, spatial and spatiotemporal exponents. 
Comoving exponents are reviewed in Sec. V within the framework introduced in
this paper. Some conclusive remarks are finally reported 
in Sec. VI.

\section{Entropy potential}

The simplest context, where a discussion on the entropy potential can be
set in, is provided by the linear diffusion equation for the field $u(x,t)$
\be
\label{diffusion}
\partial_t u = \gamma u + D \partial_x^2 u \; ,
\ee
which can be interpreted as the linearization of (the scalar version of)
Eq.~(\ref{pde}) around a uniform stationary solution
$y(x,t) = \hbox{const}$. The linear stability analysis amounts to assuming
a perturbation of the form
\be
\label{dist}
 u(x,t) \sim\exp \left({\tilde \mu} x + {\tilde \lambda}t\right) \;,
\ee
where ${\tilde \lambda} = \lambda + i \omega$ and ${\tilde \mu} = \mu + i k$
are complex numbers the real parts of which denote temporal and spatial
Lyapunov exponents, respectively, while the imaginary parts represent the
integrated densities of spatial ($\omega$) and temporal ($k$) exponents.
Substituting Eq.~({\ref{dist}) in Eq.~(\ref{diffusion}) we obtain
\be
\label{cmplx}
\tilde \lambda = \gamma + D \tilde \mu^2  \; .
\ee
By separating real and imaginary parts, we get
\bey
\label{lamu}
&&\lambda(\omega,k) = \gamma + \left(\omega^2- 4 D^2k^4\right)/(4Dk^2) \\
&&\mu (\omega,k) = \omega/(2Dk) \quad . \nonumber
\eey
As already discussed in LPT, $\omega$ and $k$ play the same role as the 
integrated densities $n_\mu$ and $n_\lambda$ and can be explicitely obtained 
by inverting Eqs.~(\ref{lamu}),
\bey
\label{nn}
&&n_\lambda \equiv k = \sqrt{\mu^2- {\lambda-\gamma \over D}} \\
&&n_\mu \equiv -\omega = - 2D\mu\sqrt {\mu^2 - {\lambda-\gamma \over D}}
 \quad . \nonumber
\eey
The minus sign in the definition of $n_\mu$ is just a matter of convention: 
we adopt this choice for consistency reasons with LPT.\par
The above sets of equations (\ref{lamu},\ref{nn}) stress, in a particular
instance, the general observation reported in LPT that either the pair 
$(n_\mu,n_\lambda)$ or $(\mu,\lambda)$ suffices to identify a given 
perturbation, the remaining two variables being determined from the Lyapunov 
spectra. However,
any two items in the set $\{\mu,\lambda,n_\mu,n_\lambda\}$ can be chosen to
be the independent variables. The above two choices are preferable for
symmetry reasons; however, the pairs $(\mu,n_\lambda)$ and
$(\lambda,n_\mu)$ turn out to be the best ones for the identification of
a single function, the entropy potential, which determines all stability
properties.\par
In fact, as it is clear from Eq.~(\ref{cmplx}), we can condense the two
real functions needed for a complete characerization of the stability
properties into a single analytic complex expression. Now, the mere
circumstance that $\lambda(\mu,n_\lambda)$ and $n_\mu(\mu,n_\lambda)$ are
the real and imaginary parts of the analytic function
$\tilde \lambda^*(\tilde \mu)$ has an immediate and important 
consequence:\footnote{The reference to the complex conjugate
variable again follows from the convention adopted for $n_\mu$} 
Cauchy-Riemann conditions are satisfied and it is possible to write
$\lambda$ and $n_\mu$ as partial derivatives of the same real function,
\bey
\label{enpot1}
  &&{ \partial \Psi \over \partial n_\lambda} = \lambda \\
  &&{ \partial \Psi \over \partial \mu} = -n_\mu \quad ,\nonumber
\eey
where $\Psi$ is the imaginary part of the formal integral $\tilde \Psi$ of
$\tilde \lambda$ with respect to $\tilde \mu$. Equivalently, one might call 
into play the real part of $\tilde \Psi$, as it is known that the latter 
contains the same amount of information.\par

In the case under investigation, we find
\be
\Psi(\mu,n_\lambda) =  n_\lambda\left(\gamma + D\mu^2\right)
- {D\over 3}n_\lambda^3 \; ,
\ee
which, together with Eq.~(\ref{enpot1}), provides a complete characterization
of the system.\par

Another, less trivial, example where the linearized problem leads to an 
analytic function for the eigenvalues is the the 1D complex Ginzburg-Landau 
equation \cite{mannev,croho} 
\be
\label{cgle}
\partial_t A  = (1+i c_1) \partial_x^2 A +A -(1-ic_3) A|A|^2  \; ,
\ee
where $A(x,t)$ is a complex field and $c_1$ and $c_3$ are
real positive parameters. The stability of the ``phase winding'' solutions
$A(x,t)= A_0 \exp(i (\nu x - \omega_0 t))$, with $A_0 = \sqrt{1-\nu^2}$
and $\omega_0 = -c_3 + (c_1 +c_3) \nu^2$, are ruled by the following 
equation for the (complex) perturbation $u(x,t)$
\be
\label{lincgle}
\partial_t u = (1+i c_1) (\partial_x^2 u  + 2 i \partial_x u)
- (1-i c_3)(1-\nu^2)(u+u^*) \quad ,
\ee
together with its complex conjugate for $u^*$ considered as an independent 
variable. The eigenvalue problem is solved assuming again
\be
\label{solcgle}
u(x,t)=u_0\exp \left({\tilde \mu} x + {\tilde \lambda}t\right) \quad ;
\quad
u^*(x,t)=u_0^*\exp \left({\tilde \mu} x + {\tilde \lambda}t\right)\quad, 
\ee
and equating to zero the determinant of the resulting linear system, to get 
the analytic (implicit) relation between $\tilde \lambda$ and $\tilde \mu$
\be
\label{eigen}
\left(\tilde \lambda + 1 - \nu^2 - {\tilde \mu}^2 + 2 c_1 \nu {\tilde \mu}\right)^2 
+ \left(c_1 {\tilde \mu}^2 + 2 \nu {\tilde \mu} + (1-\nu^2)c_3 \right)^2 = 
(1 +c_3^2)(1-\nu^2)^2,
\ee
which is analogous to Eq.~(\ref{cmplx}) 

On the basis of the examples discussed here and in the Appendix, 
one can convince himself that the analyticity property of the eigenvalue 
equation appears to be very general. Periodicity in time simply 
leads to multiply several r.h.s.'s all depending on $\tilde \mu$, while 
periodicity in space requires distinguishing between different sites on the 
lattice. In the latter case, the equivalent of Eq.~(\ref{cmplx2}) is 
obtained by equating to zero a suitable determinant, where the only variable 
is $\tilde \mu$. There is no reason to expect that different conditions 
should hold in aperiodic regimes.\par
The above approach is, in some sense, a
generalization of dispersion relations which are normally introduced for
the characterization of elliptic equations. In that case, the only
acceptable linear solutions are propagating plane waves, that is
$\lambda=\mu=0$ for (almost) all wavenumbers. From our point of view, this
implies a strong simplification since the mutual relationships among 
$\{n_\lambda,n_\mu,\lambda,\mu\}$ reduce to the link between spatial
and temporal wavenumbers. Moreover, it is obvious
that, because of the degeneracy, not all representations are equivalent
(in particular, the $(\mu,\lambda)$ plane is totally useless).\par
We must stress that the methodology that we are trying to develop in these
two papers applies to general systems where propagation coexists with
amplification (or damping). This is by no means a limitation, as all
models introduced for the characterization of space-time chaos are in this
class.\par
However, the most serious obstacle to a rigorous proof of the general
validity of Eq.~(\ref{enpot1}) is represented by the identification of
the integrated densities $n_\mu$, $n_\lambda$ with the wavenumbers $\omega$ and
$\kappa$, respectively. In the presence of spatial disorder, the Lyapunov
vectors are no longer Fourier modes: one can at most determine an average
wavenumber by counting the number of nodes in the eigenfunctions. This is
not a problem in the absence of temporal disorder, when the node theorem
applies \cite{mattis}. However, in more general cases, it is no longer 
possible to speak of eigenfunctions and we are not aware of any generalization 
to overcome the difficulty. For such a reason, we have performed some direct 
numerical check to verify the correctness of our conjectures. \par
Before discussing numerical simulations, let us come back to the problem of
the representation. In the above part, we have seen that the choice of the 
pair of independent variables $(\mu,n_\lambda)$ was very fruitful for the
identification of a potential. However, the asymmetry of such a choice calls
for transferring the above result in either representation
proposed in LPT. This step can be easily done with the help of Legendre
transforms. We discuss the transformation to the plane $(\mu,\lambda)$, any
other transformation being a straightforward generalization of the same
procedure.\par
>From the first of Eq.~(\ref{enpot1}), we see that $\lambda$ and $n_\lambda$
can, indeed, be considered as conjugate variables in a Legendre transform
involving $\Psi$.  The conjugate potential is naturally
\be
  \Phi \equiv \lambda n_\lambda - \Psi \; .
\ee
It is easily seen that in the new representation, the following relations
hold
\bey
\label{enpot2}
  &&\partial_\lambda \Phi  = n_\lambda \\
  &&\partial_\mu \Phi  = n_\mu \nonumber \quad .
\eey
Accordingly, the potential $\Phi$ is the appropriate function which
allows determining the two integrated densities in the symmetric 
representation $(\lambda,\mu)$. We call $\Phi$ the entropy potential since it
coincides with the Kolmogorov-Sinai entropy density along a suitable line 
(see Sec. III).\par
In Fig.~\ref{gillette}, we present a numerical reconstruction of
$\Phi(\mu,\lambda)$ in terms of its contour levels for the homogenoeus
CML, namely model (\ref{mappa}) with $f(x)=rx({\rm mod \quad 1})$. 
The entropy potential is obviously known up to an additive arbitrary 
constant that we have fixed by imposing that the value attained on the 
upper border is equal to zero. The potential increaeses monotonously from 
top to bottom. It is clear that outside the allowed region delimited by the 
solid curves, $\Phi$ is a linear function of $\mu$ and $\lambda$.\par
The structure of the potential does not substantially change for more general
CMLs. 
We have tested Eq.~(\ref{enpot2}) for a lattice of logistic maps 
($f(x)=4x(1-x)$) with $\varepsilon=1/3$ by integrating along two different 
paths in the $(\mu,\lambda)$ plane (see Table 1). The difference is so small 
that we can confirm that the relations are valid, within the numerical error. 

\section{Spatiotemporal exponents}

In the perspective of a complete characterization of space-time chaos, one 
should consider the possibility of viewing a generic pattern as being
generated along directions other than time and space axes. In fact, once a
pattern is given, any direction can, a priori, be considered as an
appropriate ``time'' axis. Accordingly, questions can be addressed about the
statistical properties of the pattern when viewed in that way.

\subsection{Definitions}

For the moment, we assume that the pattern is continuous both along space
and time directions; we shall discuss later how the definitions can be extended
to CML models. Let us consider a given spatiotemporal configuration of the 
field ${\bf y}(x,t)$, generated, say, by integrating Eq.~(\ref{pde}). 
When arbitrary directions are considered in the $(x,t)$ plane, the coordinates
must be properly scaled in order to force them to have the same dimension. We 
choose to multiply the time variable by $c$, where $c$ is a suitable constant 
with the dimension of a velocity.  Moreover, let $\vartheta$ denote the 
rotation angle of the tilted frame $(x',ct')$ with respect to the initial one 
$(x,ct)$, adopting the convention that positive angles correspond to clockwise 
rotations. Sometimes, it will be more convenient to identify the new frame by 
referring to the velocity $v = c\tan\vartheta$. The limit cases $v=0$ 
($\vartheta=0$) and $v=+\infty$ ($\vartheta=\pi/2$) correspond to purely
temporal and purely spatial propagations, respectively. 
The coordinate transformation reads as
\bey \label{trasfo}
&&ct' = \beta \left(ct +{v\over c} x \right) \\ 
&&x' = \beta \left( -vt + x \right) \quad , \nonumber
\eey 
where $\beta \equiv 1/\sqrt{(1+v^2/c^2)}$. The physical meaning of $v$ is
transparent: it can be interpreted as the velocity in the old 
frame of a point stationary in the tilted frame (constant $x'$).

The new field ${\bf y}(x',ct')$ can be thought of as being the result of the
integration of the model derived from the original one after the change of 
variables (\ref{trasfo}). Although it is not obvious whether the invariant 
measure in the initial frame is still attracting in the new frame 
(see Ref.~\cite{bidime} for a discussion of this point), one can anyhow study 
the stability properties by linearizing and defining the Lyapunov exponents in 
the usual way. 

In CML models, the discreteness of both the space and the time lattice 
leads to some difficulties in the practical construction of tilted frames. In 
fact, only rational values of the velocity $v$ can be realized in finite
lattices (in this case, it is natural to assume that the lattice spacing is 
the ``same'' along the spatial and the temporal directions and, accordingly,
to set $c=1$). Moreover, writing the explicit expression of the model 
requires introducing different site types. For this reason, we discuss in the 
following the simplest nontrivial case $v=1/2$, the generalization to other
rational velocities being conceptually straightforward.

A generic spatial configuration in the tilted frame is defined by sites of the 
spatiotemporal lattice $(i,n)$ connected by alternating horizontal (as in the 
usual case) and diagonal bonds (see Fig.~\ref{tilt}). By suitably adjusting 
the relative fraction of the two types of links all rotations between 0 and 
$\pi/4$ can be reproduced.
The explicit expression of the updating rule requires a proper numbering of 
the consecutive sites. Moreover, as seen in Fig.~\ref{tilt}, it involves the 
``memory'' of two previous states. 

Finally, an exact implementation of the mapping rule requires acausal boundary
conditions, since the knowledge of future (in the original frame) states is
required \cite{bidime} (this is a general problem occurring also in the 
continuous case). As we are interested in the thermodynamic limit, we bypass
the problem by choosing periodic boundary conditions. Such a choice has been 
shown not to affect the bulk properties of the dynamical evolution 
\cite{bidime}.\par
In the updating procedure, two different cases are recognized: the variable
$y$ is either determined from the past values in the neighbouring sites, or
it requires the newly updated $y$-value on the right neighbour (see
Fig.~\ref{tilt}). For $v=1/2$, this can be done by simply distinguishing
between even and odd sites,
\begin{eqnarray}
\label{v2}
&&X^{2i}_{n+1} 
   = f\left((1-\varepsilon)X^{2i}_n+ {\varepsilon \over 2}
      \left[  Y^{2i-1}_n+X^{2i+1}_n \right]\right)\\ 
&&X^{2i+1}_{n+1} 
   = f\left((1-\varepsilon)X^{2i+1}_n+ {\varepsilon \over 2}
      \left[  X^{2i}_n +X^{2i+2}_{n+1}\right]\right)\quad ,\nonumber 
\end {eqnarray}
where $i=1,\ldots,L/2$ ($L$ is assumed to be even for simplicity), while
\be
Y^j_{n+1} \equiv X^j_n \quad ,
\ee
are additional variables introduced to account for the dependence at time
$n-1$. Taking into account that $X_{n+1}^{2i+2}$ can be determined from
the $X$ and $Y$ variables at time $n$, the mapping can be finally expressed
in the usual synchronous form $(X_n^i,Y_n^i)\to (X_{n+1}^i,Y_{n+1}^i)$, but
with an asymmetric spatial coupling with next and next-to-next nearest
neighbours. The Lyapunov exponents $\eta_j$ can now be computed with the 
usual technique \cite{benettin}.

In analogy with the original model, we expect again that, in the limit of 
infinitely extended systems, the set of exponents $\eta_j(v)$ will converge 
to an asymptotic form,
\be
 \eta_j(v) \to \eta(v,n_\eta)\quad, 
\ee
where $n_\eta$ is the corresponding integrated density. We will refer to 
this function as the spatiotemporal Lyapunov spectrum (STLS). In the limit 
cases $v=0,+\infty$ ($\vartheta=0,\pi/2$), the STLS reduces to the 
standard temporal and spatial spectrum, respectively.

The recursive scheme (\ref{v2}) implies an increase of the phase-space 
dimension by a factor (1+1/2) (in general $1+v$). Actually, as
we will argue, these new degrees of freedom are not physically relevant.
However, for consistency reasons with the original rescaling of the spatial 
variable, we choose to normalize the spatiotemporal density between 0 and 
$1+v$ (the time units are, instead, left unchanged by the above construction).

The generalization to asymmetric maps (\ref{amappa}) is straightforward: it
removes the degeneracy $v \to -v$. Numerical results for logistic maps,
indicate that the dependence of the positive exponents on the velocity is
quite weak in the fully symmetric case $\alpha =1/2$ (for instance, the
maximum exponent exhibits a 20\% variation in the whole $v$ range), while
it is remarkable for asymmetric couplings. In every case, the negative part
of the spectrum sharply changes with the velocity. This is consistent with
the results obtained for delayed maps in Ref.~\cite{bidime}.

\subsection{Representation in the $(\mu,\lambda)$ plane}

Spatiotemporal exponents can be put in relation with $\mu$ and $\lambda$ 
by rewriting the general expression for a perturbation in a frame rotated by
an angle $\vartheta$,
\be
\label{rota}
   \exp(\mu x+\lambda t)=\exp(\mu'x'+\lambda' t')\quad .
\ee
Such an equation induces a rotation of the same angle in the
$(c\mu,\lambda)$ variables,
\bey
\label{trasfolamu}
&&\lambda' = \beta \left( \lambda +v \mu \right) \\ 
&&\mu' = \beta \left( -(v/c^2)\lambda + \mu \right) \quad . \nonumber
\eey 
The above equations allow studying the stability with respect to generic
perturbations with an exponential profile along $x'$. For simplicity, we 
shall consider only uniform perturbations, 
\be
\label{eta}
   \exp(\mu x+\lambda t)=\exp(\eta t') \quad ,
\ee
where the growth rate $\eta$ denotes the spatiotemporal exponent. Notice that
we have changed notations from $\lambda'$ to $\eta$, to understand that the 
condition $\mu' = 0$ is fulfilled. From the second of Eq.~(\ref{trasfolamu}),
uniform perturbations in the rotated frame correspond to points along the line
$\cal L$ 
\be
\label{velvel}
\lambda = c^2\mu/v  \quad ,
\ee
in the $(\mu,\lambda)$ plane.

Whenever the evolution of an exponentially localized perturbation of type
(\ref{rota}) is considered, it is natural to introduce the quantity
$\hat V=\lambda/\mu$, which can be interpreted as the velocity of the front
\cite{disturbi}. Eq.~(\ref{velvel}) connects this velocity with that of the
rotated frame, 
\be
\hat V =c^2/v \quad .
\ee
Therefore, on the basis of definition (\ref{eta}), $(x',ct')$ can be
interpreted as the reference frame in which the front associated with the
perturbation propagates with an ``infinite'' velocity.

The explicit expression for $\eta$ is
\be
\label{distan}
\eta = \sqrt{\lambda^2 + (c \mu)^2} = \lambda/\beta \quad .
\ee
Such a relation can be turned into a self-consistent equation for the
maximum Lyapunov exponent $\eta_{\rm max}$ by imposing the constraint that
the pair $(\mu,\lambda)$ lies on the line $\lambda = \lambda_{\rm max}(\mu)$, 
namely 
\be
\label{selfcon}
 \eta_{\rm max} = {1 \over \beta} \lambda_{\rm max} \left({v\over c^2} 
   \beta\eta_{\rm max} \right)\quad .
\ee 
Some ambiguities arise when velocities $v>c$ are considered, since the line
$\cal L$ intersecates $\lambda_{\rm max}(\mu)$ in two points as seen in  
Fig.~\ref{bordi}. This phenomenon was already noticed in Ref.~\cite{disturbi},
while discussing the propagation of exponentially localized disturbances in 
the original reference frame. Moreover, it has been shown that only the front
corresponding to the smaller value of $\mu$ is stable, except for some cases
where a nonlinear mechanism intervenes dominating the propagation process
\cite{nonlin}.

At $v = c^2/V_*$ the two intersections degenerate into a single tangency
point. This condition defines $V_*$, which can be interpreted as 
the slowest propagation velocity of initially localized disturbances 
\cite{disturbi}.

The extension of Eq.~(\ref{selfcon}) to the rest of the spectrum requires
to connect $n_\lambda$ and $n_\mu$ with $n_\eta$. In the next section, we
will show how to perform such a step with the help of the entropy potential.
Here, we limit ourselves to discuss the structure of the STLS for different
values of the tilting angle $\vartheta$. In Fig.~\ref{ruotbd}, we report the borders
of the bands, which can be determined from the intersections of $\cal L$ with
the border $\partial {\cal D}$ of the domain of allowed perturbations
(see Fig.~1 of LPT and Fig.~\ref{bordi}). For $\vartheta=0$ (temporal case) 
a single band is present but, as soon as $\vartheta >0$, a second negative band 
arises from the intersections with the branch diverging to $-\infty$ at $\mu=-\mu_c$.
For $\vartheta > \pi/4$, the negative band disappears and a positive band
arises from the intersections with the branch diverging to $+\infty$ with
slope $v=c$. A single band spectrum is again recovered for $\vartheta \ge
\vartheta_* = {\rm atan}(c/V_*)$.

Notice that in symplectic maps, the STLS is symmetric for any value of 
$\vartheta$ (see LPT) so that positive and negative bands appear and disappear
simultaneously.

It is worthwhile to illustrate some of the above considerations in the
simple case of the linear diffusion equation (\ref{diffusion}). The
expression for $\lambda_{\rm max}(\mu)$ can be obtained from
Eq.~(\ref{cmplx}), by setting $k$ and $\omega$ equal to 0. Accordingly,
Eq.~(\ref{selfcon}) reads as
\be
\label{etamax}
\beta\eta_{\rm max} = \gamma+D\left({v\over c^2}\beta\eta_{\rm max}
\right)^2  \quad .
\ee
On the other hand, the model equation in the rotated frame can be obtained
from the substitutions
\bey
\label{trasdif}
&&\partial_t \to  \beta \left(\partial_{t'} -v\partial_{x'} \right) \\ 
&&\partial_x \to \beta \left( {v\over c^2}\partial_{t'} + \partial_{x'}
\right) \quad . \nonumber
\eey
By introducing the usual Ansatz for the shape of the perturbation,
\be
\label{uprimed}
u(x',t') \sim  \exp \left[i\kappa x' + (\eta+i\Omega)t'\right] \quad ,
\ee
separating the real from the imaginary part, and eliminating
$\Omega$, we obtain the integrated density of spatiotemporal exponents
\be
\label{rls}
\kappa(\eta,v) = \beta\left( 1-2D{v^2\over c^4}\beta\eta\right)
\sqrt{{v^2\over c^4}(\beta\eta)^2 - { \beta\eta \over D}
+ {\gamma\over D}} 
\ee
and the corresponding STLS $\eta(\kappa,v)$. Dimensional analysis shows that
$\kappa$ is an inverse length, as expected for a density of exponents.
Notice that Eq.~(\ref{etamax}) is recovered, by setting $\kappa = \Omega = 0$
in Eq.~(\ref{rls}). 

In this and more general continuous models, we should remark that the line
$\cal L$ intersects $\lambda_{\rm max}(\mu)$ twice for any arbitrarily
small $v$. This is because the Laplacian operator sets no upper limit to the 
propagation velocity of disturbances.

\section{From the entropy potential to dynamical invariants}

The present section is devoted to establish the consequences of the
existence of the entropy potential on the Lyapunov spectra and other
dynamical indicators such as the Kolmogorov-Sinai entropy and the
Kaplan-Yorke dimension of the attractor. In order to keep the notations as
simple as possible, we assume that time and space coordinates are scaled in
such a way that $c=1$.

\subsection{Spatiotemporal exponents} 
 
The very existence of the entropy potential $\Phi$ implies that the Lyapunov
spectrum in a frame tilted at an angle $\vartheta$ (recall that $\vartheta$ is
the angle from the $\lambda$-axis) can be obtained by computing the
derivative of $\Phi$ along the direction
$\vec u = (\sin \vartheta, \cos \vartheta)$ in the $(\mu,\lambda)$ plane. In
fact, this is a straightforward generalization of the previous findings that
$n_\mu$ and $n_\lambda$ are the derivatives of $\Phi$ along the $\mu$ and
$\lambda$ direction, respectively. Accordingly, the STLS is linked to the
TLS and SLS by the following general equation
\be
\label{formula}
n_\eta(v,\eta) = \vec u \cdot \nabla \Phi =
\beta \left [ v n_\mu + n_\lambda \right ] \quad ,
\ee
where $\nabla=(\partial_\mu,\partial_\lambda)$ is the gradient in the
$(\mu,\lambda)$ plane, and the r.h.s of the above formula is evaluated for
\bey
&&\mu = v \beta\eta \\
&&\lambda = \beta\eta \quad .
\eey
Such a relation can be directly verified for the diffusion equation from
Eqs.~(\ref{nn}),(\ref{rls}). Further, more significative tests have been
performed by checking numerically the validity of Eq.~(\ref{formula}) in some 
lattice models involving, e.g., logistic and homogeneous chains, the spectra
of which are reported in see Fig.~\ref{spetr} (notice that, since the 
time axis has not been renormalized in the tilted frame, the factor $\beta$ 
need not be introduced).

\subsection{Entropy}

Kolmogorov-Sinai entropy $H_{KS}$ is a measure of the information-production
rate during a chaotic evolution. An estimate of $H_{KS}$ is given by the
Pesin formula \cite{eck} as the sum $H_\lambda$ of the positive Lyapunov
exponents. While it is rigorously proven that $H_{KS} \le H_\lambda$,
numerical simulations indicate that, in general, an equality holds.
In spatially extended systems, $H_{KS}$ is believed to be proportional to
the system size \cite{grass}. For this reason, it is convenient to introduce
the entropy density $h_\lambda$ which, in the thermodynamic limit, is
computed as the integral of the positive part of the Lyapunov spectrum.

Therefore, it is natural to extend the definition of entropy to tilted
frames as an integral along the line $\cal L$,
\be
\label{pesin}
h_\eta =  \int_{n_{\rm min}}^{n_{\rm max}} \eta (n_\mu,n_\lambda) dn_\eta
\quad ,
\ee
where $n_{\rm min}$ is the integrated density in the point where the line
$\cal L$ intersects $\partial {\cal D}$, i.e. where $\eta = \eta_{\rm max}$,
while $n_{\rm max}$ is measured in the origin, i.e. where $\eta = 0$. In the
limit $v = 0$, the above equation reduces to the definition of the
density $h_\lambda$, which refers to the original reference frame. For
$v \to \infty$, instead, we obtain the ``spatial'' entropy density $h_\mu$.

Numerical simulations performed with different CML models indicate that 
$h_\lambda < h_\mu$. This can be explained by the following argument. The 
patterns obtained asymptotically by iterating the model in the original
reference frame are, in general, unstable if generated along the spatial
direction \cite{Torcini}. In other words, the spatiotemporal attractor is a
(strange) repellor of the spatial dynamics. Accordingly, part of the local
instability accounted for by the sum of positive spatial Lyapunov exponents
is turned into a contribution to the escape rate from the repellor
\cite{Tel} and $h_\mu$ must be larger than the entropy $h_\lambda$ of the
original pattern. In continuous models this inequality is brought to the 
extreme case, as $h_\mu$ is infinite.

Numerical simulations of a lattice of logistic maps with $0<v<1/V_*$,
confirm the previous evidence \cite{bidime} that $h_\eta$ is independent of
$v$. Moreover, it is found that $h_\eta$ is constant and equal to $h_\mu$ for
$v$ large enough. Both results have a simple explanation in terms of the
entropy potential. The very definition of $h_\eta$ suggests that it is more
convenient to refer to the $(n_\mu,n_\lambda)$ plane. Indeed, as long as the
outer intersection of $\cal L$ with $\partial {\cal D}$ in the plane
$(\mu,\lambda)$ occurs on the same branch, the endpoints of the integrals
(\ref{pesin}) are the same (in fact, in Fig.~3 of LPT it is shown that each 
connected component of $\partial {\cal D}$ is mapped onto a single point in the
$(n_\mu,n_\lambda)$ plane).

If we now notice that the integrand in Eq.~(\ref{pesin}) is the gradient of
the potential
\be
\label{conphi}
\tilde \Phi = \lambda n_\lambda + \mu n_\mu - \Phi \quad ,
\ee
it is mathematically obvious why $h_\eta$ is independent of $v$ in suitable
intervals.

Another explanation of the above result can be found by referring directly
to the plane $(\mu,\lambda)$. Integrating by parts, Eq.~(\ref{pesin}) can be
rewritten as
\be
\label{pesin2}
h_\eta =  n_{\rm min} \eta_{\rm max} - \int_0^{\eta_{\rm max}}
  n_\eta (\mu,\lambda) d \eta \quad .
\ee
As $n_{\rm min}=0$ along the upper border (see LPT), we can compute 
$h_\eta(v)$ for $v<1$ by integrating the gradient of $\Phi$ along the line 
$\cal L$ from $(0,0)$ to the upper border itself. Since the upper
border is an equipotential line (see also Fig.~\ref{gillette}), $h_\eta$ is
independent of $v$. The same argument can be repeated for $v>1/V^*$
by suitably shifting $n_\mu$ and $n_\lambda$.

The independency of $h_\eta$ of $v$ has also a physical interpretation.
The Kolmogorov-Sinai entropy density, in fact, is the amount of information
needed to characterize a space-time pattern (apart from the information flow
through the boundaries \cite{grass}) divided by its temporal duration
and the spatial extension, i.e. divided by the area. Therefore, $h_{KS}$ is
independent of they way the axes are oriented in the plane, i.e. of
the velocity $v$. As a consequence, $h_\eta = h_{KS}$ for all $v<1$.

The above conclusion still holds when the STLS exhibits a positive band as well
(which is always the case in continuous models), provided that the content of 
such a band is discarded. Accordingly, we can conclude that the new degrees
of freedom, associated to the positive band, which appear in the rotated frame
are just physically irrelevant directions which turn the original attractor
into a repellor. If $v> c^2/V_*$, the two bands merge together and it is not 
anymore possible to distinguish between unstable but irrelevant directions and 
the unstable manifold of the original attractor. Presumably, this means that 
the repellor is turned into a strange repellor with a singular measure along 
some (all) unstable directions.

\subsection{Dimension}

A second important indicator of the ``complexity'' of a spatiotemporal dynamics
is the fractal dimension. An upper bound $D_{KY}$ to it is given by the
Kaplan-Yorke formula \cite{eck}. The existence of a limit Lyapunov spectrum,
implies that $D_{KY}$ is proportional to the system size \cite{grass}, so that 
it is convenient to introduce the dimension density $d_{KY}$. In the framework 
of the present paper, it is natural to extend the definition to generic 
velocities. 
The dimension density satisfies the integral equation,
\be
\label{ky}
\int _0^{d_{KY}} \eta(v,n_\eta) dn_\eta = 0 \quad .
\ee
As for the entropy density, Eq.~(\ref{ky}) can be more easily interpreted with
reference to the $(n_\mu,n_\lambda)$ plane. In fact, the curve implicitely
defined by the above constraint is the equipotential line $\cal C$ 
\be
  \tilde \Phi (n_\mu,n_\lambda) = 0  \quad .
\ee
The dimension density $d_{KY}(v)$ can, in turn, be determined from 
Eq.~(\ref{formula}) at the intersection point between $\cal C$ and the
image of $\cal L$ in the plane $(n_\mu,n_\lambda)$. 

At variance with the entropy density, $d_{KY}(v)$ changes with $v$ (see 
Fig.~\ref{dim}) even if we avoid considering the second positive band.
In fact, while $h_\eta$ is an information divided by a space-time area, 
$d_{KY}(v)$ is a number of degrees of freedom divided by a length, measured 
orthogonally to the propagation axis. Thus, at least from a dimensional 
point of view, it is meaningless to compare $d_{KY}(v)$ for different 
velocities. However, one can reduce temporal to spatial lengths by
introducing the scaling factor $c$ and, in turn, ask himself how the
dimension changes with $c$. It is easily seen that the scaling dependence
on $c$ is expressed by the following relation,
\be
\label{scale}
d_{KY}(v,c_1) \sqrt{1 + \left ({ v \over c_1 } \right )^2} = 
d_{KY}(v,c_2) \sqrt{1 + \left ({ v \over c_2 } \right )^2} \quad , 
\ee
The (completely arbitrary) choice of $c$ reflects in different dependences of 
$d_{KY}$ on $v$. A natural procedure to fix $c$ is by minimizing the 
dependence of $d_{KY}$ on the observation angle. This amounts to choosing the 
time units in such a way as to make the 2D pattern as isotropic as 
possible. In homogeneous CMLs, the procedure is so effective that a suitable 
choice of $c$ allows removing alomost completely the velocity dependence as 
seen in Fig.~\ref{dim}a, where the results for the natural value $c=1$ are 
compared with those for $c=3$.\par
More in general, however, it is not possible to achieve such a complete
success. This is, for instance, the case of the logistic CML, where the 
dimension drop for $c=1$ is too large to be compensated by any choice of $c$ 
(see Fig.~\ref{dim}b, where the curve for $c=1$ is compared with the best
results obtained for $c=+\infty$).

A further indicator which is sometimes useful in characterizing the chaoticity
of a given extended system is the dimension density $d_u$ of the unstable 
manifold. This dimension is nothing but $n_\eta$ in the point where $\eta=0$,
i.e. in the origin, and its expression simply reads as
\be
d_u = \beta n_\lambda(0,0) \quad .
\ee
being $n_\mu(0,0) \equiv 0$. The choice $c=+\infty$ of the scaling factor 
removes exactly the dependence on the orientation of the reference frame.
This choice is equivalent to measuring lengths in the untilted frame.

\section{Comoving exponents}

Another class of indicators, introduced to describe convective instabilities
in open-flow systems, consists of the so-called comoving or
velocity-dependent Lyapunov exponents \cite{como}. They quantify the 
growth rate of a localized disturbance in a reference frame moving with 
constant velocity $V$. Given an initial perturbation $u(x,0)$ which is
different from zero only within the spatial interval $[-L_0/2,L_0/2]$, 
numerical analyses indicate 
\be
\label{delta}
   u(x,t) \sim \exp\left(\Lambda(x/t)t\right) \quad ,
\ee
for $t$ sufficiently large. Eq.~(\ref{delta}) defines the comoving Lyapunov 
exponent $\Lambda$ as a function of $V=x/t$. The initial width $L_0$ of the 
disturbance is not a relevant parameter, since a generic perturbation grows 
with the maximum rate.\footnote{In the particular case of a $\delta$-like 
initial profile, the definition of local Lyapunov exponent introduced in 
Ref.~\cite{pikov} is recovered.}

The definition of $\Lambda$ can be extended to a whole spectrum of comoving
exponents by looking not just at the local amplitude of the perturbation but
also at its shape \cite{Mayer}. Since the physical meaning of the rest of the
spectrum is still questionable, in the following we limit ourselves to discuss
the maximum. 

As a matter of fact, the limit $t \to \infty$ (required by a meaningful
definition of an asymptotic rate) implies the infinite-size limit.
Therefore, one must carefully keep under control the system size, when
longer times are considered. This is perhaps the most severe limitation
against an accurate direct measurement of $\Lambda$.

It can be easily shown that $\Lambda(V)$ is connected with the maximal 
temporal Lyapunov exponent $\lambda_{\rm max}(\mu)$ by a Legendre-type 
transformation \cite{Torcini}. Eq.~(\ref{delta}) implies that the perturbation
has a locally exponential profile with a rate 
\be
\label{leg1}
\mu = {d \Lambda(V) \over d V} \quad 
\ee
in the point $x = Vt$. On the other hand, we know that such a profile evolves 
as 
\be
\label{delta2}
u(Vt,t) \sim \exp [ (\lambda_{\rm max}(\mu) + \mu V)t] \quad .
\ee
By combining Eqs.~(\ref{delta}) and (\ref{delta2}), we obtain
\be
\label{leg2}
   \Lambda(V) = \lambda_{\rm max}(\mu) + \mu {{ d \lambda_{\rm max}(\mu)}
  \over d \mu} \quad , 
\ee
which, together with Eq.~(\ref{leg1}) can be interpreted as a Legendre
transform from the pair $(\Lambda,V)$ to the pair $(\lambda_{\rm max},\mu)$. 
The inverse transform reveals the further constraint
\be
V = {{ d \lambda_{\rm max}(\mu)} \over d \mu} \; .
\ee
Eq.~(\ref{leg2}) states that $\Lambda(V)$ is the growth rate of an 
exponentially localized perturbation with a given $\mu$ value as determined 
from the condition Eq.~(\ref{leg1}). However, the perturbation itself 
propagates with yet another velocity, $\tilde V = \lambda_{\rm max}(\mu)/\mu$. 
As a matter of fact, $\tilde V$ and $V$ correspond to phase and group 
velocities for propagating waves in linear dispersive media. In particular, 
the ``phase'' velocity $\tilde V$ can be larger than the ``light'' velocity 
($c=1$ in CML with nearest neighbour coupling), while $V$ is bounded to be 
smaller.

A simple geometrical interpretation of the above Legendre transformations can 
be given with reference to the $(\mu,\lambda)$ plane. The comoving Lyapunov
exponent $\Lambda(V)$ is the distance between the origin and the intersection 
of the $\lambda$ axis with the straight line of slope $V$, tangent to the upper 
temporal border. If the system is chaotic, such an intersection remains 
positive for $V \le V_*$. Therefore, as already remarked, $V_*$ is the maximum 
velocity of disturbance propagation. Indeed, along the worldlines with 
$V >V_*$, the disturbance does not vanish exactly, but decreases exponentially 
in time.

Whenever a Legendre transform comes into play, some attention must be payed
to the concavity of the functions involved in the transformation.
In the present context, this is the case of frozen random patterns where the
border of the allowed region exhibits a change of concavity at $\mu=\mu_1$
(see Fig.~6 of LPT). This implies that for $\mu <\mu_1$, the maximal temporal 
exponent is constant and equal to $\lambda_{\rm max}(\mu_1)$. The corresponding
``phase transition'' reflects itself as a linear dependence of the comoving 
Lyapunov exponent on the velocity for $|V| < V_1$,
\be
\label{frocom}
   \Lambda(V) = \lambda_{\rm max}(\mu_1) - \mu_1 V  \quad .
\ee
This is evident in Fig.~\ref{froz}, where the whole set of $\Lambda$
values is reported.

It is clear that comoving, spatiotemporal and temporal exponents are related
to one anothe. However, the link is not so straightforward as one might think.
Indeed, the velocity $v$ of the rotated frame where $\eta_{\rm max}$
coincides (up to a normalization factor) with $\lambda_{\rm max}$ is equal to
$c^2/\tilde V$ and thus differs from both $\tilde V$ and $V$. 

\section{Conclusions}

In the present paper we have shown that all instability properties of
1D chaotic systems can be derived from a suitable entropy potential expressed
as a function of any pair of variables in the set 
$\{\mu,\lambda,n_\mu,n_\lambda\}$. The most appropriate representation depends
on the problem under investigation. For 
instance, the properties of Kolmogorov-Sinai entropy are more naturally 
described with reference to $(n_\mu,n_\lambda)$. This is analogous to
standard thermodynamics, where several potentials (Gibbs, Helmholtz, etc.)
are introduced to cope with different physical conditions.
 
The very notion of entropy potential implies general relations among the
classes of Lyapunov exponents introduced and discussed here and in LPT, 
namely spatial, temporal, spatiotemporal and comoving exponents. Another
remarkable consequence of the existence of an entropy potential is the
independency of the Kolmogorov-Sinai entropy density $h_{KS}$ (as determined 
from the
the spatiotemporal spectrum) of the propagation direction in the space-time
plane. Accordingly, $h_{KS}$ can be considered as a super-invariant dynamical
indicator. This is not the case of the fractal dimension, the dependence
of which provides information about the anisotropy of the pattern.

We should, however, point out that our statements are not rigorously
proved (except for some simple test models). However, since our
numerical simulations suggest their general validity, we strongly believe
that systematic analytical investigations should eventually succeed in
proving their validity. A final remark concerns the space dimensionality. 
The existence of the entropy potential stems from the analyticity of the
complex dispersion relations which, in turn, is peculiar of 1D systems. 

\acknowledgments

We thank P. Grassberger, H. Kantz, and A. Pikovsky for useful 
discussions. We also acknowledge the hospitality of ISI-Torino during the
activity of the EC Network CHRX-CT94-0546 and the program on nonlinear 
dynamics of Laboratorio FORUM-INFM.
A.T. gratefully acknowledges the European Community for the research 
fellowship No ERBCHBICT941569 and his mother-in-law G. Frese for having 
nicely taken care of him during the tremendous winter in Wuppertal.

\appendix
\section{}

The crucial point in justifying the existence of the entropy potential
is the analytic structure of the eigenvalue equation stemming from
the linearized dynamics. To support the generality of this
statement we consider in this Appendix two more examples, namely
the linear stability analysis of homogeneous solutions both of the 1D wave
equation and of CML.  

The wave equation
\be
\label{wave}
\partial_t^2 u = -m^2 u + \partial_x^2 u \\,
\ee
is the conservative analogous of Eq.~(\ref{diffusion}) ($m$ is a real 
parameter) and can be treated in a similar way, obtaining
\be
\label{linwave}
\tilde\lambda^2 = \tilde\mu^2 -m^2 \quad.
\ee 
The abovee expression justifies {\it per se} the existence of 
the entropy potential.
Incidentally, notice that the Hamiltonian nature of Eq.~(\ref{wave})
implies the degeneracy of the standard TLS in zero, since the 
uniform solution is an elliptic fixed point. The entropy potential is
determined as the real or, equivalently, the imaginary part of the
formal integral
\be
\label{integral}
\tilde\Psi(\tilde\mu)=\int\tilde\lambda d\tilde\mu =
{1\over2}\left[\tilde\mu^2-m^2\cosh^{-1}\left({\tilde\mu\over m}
\right)\right] \\.
\ee
This can be verified in the limit of a ``weak'' instability $m\to 0$, when
Eq.~(\ref{integral}) approximately reads as 
\be
\tilde\Psi(\tilde\mu)\approx
{1\over2}\left[\tilde\mu^2-m^2\log\left({\tilde\mu\over m}
\right)\right] \\.
\ee
By also expanding to the lowest order in $m$ the expressions of $\lambda$ and 
$\omega$ determined by Eq.~(\ref{linwave}), we obtain 
\bey
\label{espwave}
 &&\lambda(\mu,k) \approx |\mu|\left(
 1-{1\over2}{m^2\over \mu^2+k^2}\right) \\
 &&\omega(\mu,k) \approx  k\left(
 1+{1\over 2}{m^2\over \mu^2+k^2}\right)\nonumber \quad .
\eey
It is straightforward to verify that
\bey
\label{enreim}
  &&\partial_k {\rm Re}\tilde\Psi = 
  - \partial_\mu {\rm Im}\tilde\Psi = \omega \\ 
  &&\partial_\mu {\rm Re}\tilde\Psi = 
  \partial_k {\rm Im}\tilde\Psi = \lambda\nonumber \quad . 
\eey

For homogeneous solutions of CML models, we obtain
\be
\label{cmplx2}
 e^{\tilde \lambda} = r \left [ (1-\varepsilon) + \varepsilon \cosh\tilde 
\mu \right ] \quad,
\ee
where $r$ is the multiplier.
Unfortunately, in this case it is not possible to write down
an explicit expression for the integral $\tilde\Psi$ for generic parameter values.
We limit ourselves to discuss the problem in the limit of a small coupling,
i.e. $\varepsilon \to 0$.
Expansion of (\ref{cmplx2}) to the first order in $\varepsilon$, 
yields
\be
\label{cpot}
\tilde \Psi(\tilde \mu) \approx (\log r -\varepsilon) \tilde \mu
 + \varepsilon \sinh \tilde \mu \quad ,
\ee
and 
\bey
 &&\lambda(\mu,n_\lambda) \approx  \log r -\varepsilon 
  \left( 1-\cos k\cosh\mu \right)  \\
  &&n_\mu(\mu,n_\lambda)\approx \varepsilon \sin k \sinh\mu
  \nonumber \quad ,
\eey
which should be compared with the corresponding expressions
obtained by expanding to first order in $\varepsilon$ Eqs.~(16)
and (20) of LPT. Moreover, one can verify that the relations analogous to
Eqs.~(\ref{enreim}) hold also in the present example.

\begin{figure}
\caption[gillette]{Contour plot of the entropy potential $\Phi$ for
a homogeneous chain ($r=2$, $\varepsilon=1/3$).}
\label{gillette}
\end{figure}

\begin{figure}
\caption[tilt]{Lattice implementation of the definition of 
spatiotemporal Lyapunov exponents for $v=1/2$.}
\label{tilt}
\end{figure}

\begin{figure}
\caption[bordi]{Plot of the boundary $\partial{\cal D}$ and of the line
$\lambda= v \mu$ in the three velocity regimes for the logistic
CML $\varepsilon=1/3$. The three lines refer to the different 
cases $v < 1$ (solid), $ 1 < v < 1/V_*$ (dot-dashed) and 
$v=1/V_*$ (dashed).}
\label{bordi}
\end{figure}

\begin{figure}
\caption[ruotbd]{Boundaries of the STLS versus the tilting 
angle $\vartheta$ for the homogeneous chain ($r=2$, $\varepsilon=1/3$).}
\label{ruotbd}
\end{figure}

\begin{figure}
\caption[spetr]{Comparison between the STLS obtained by direct
numerical computation and formula (\ref{formula}) for (a) homogeneous 
($r=2$) with $v=4/5$ and (b) logistic CML with $v=3/5$ (in both 
cases $\varepsilon=1/3$).}
\label{spetr}
\end{figure}

\begin{figure}
\caption[dim]{Kaplan-Yorke dimension density $d_{KY}$ obtained from 
the STLS versus the tilting angle $\vartheta$ for (a) homogeneous
($r=1.2$) and (b) logistic CML models: in both cases 
$\varepsilon=1/3$. Circles refer to the scaling factor $c=1$, while
crosses correspond to $c=3$, $+\infty$ in (a), (b), respectively. 
}
\label{dim}
\end{figure}

\begin{figure}
\caption[froz]{Maximum comoving Lyapunov exponent $\Lambda(V)$ 
for a  frozen random pattern obtained as  Legendre transform versus 
$V$ (for comparison see also Fig. 6 in LPT). The vertical line indicates
the position of the critical velocity $V_1$ (see the text for definition).}
\label{froz}
\end{figure}

\begin{table}
\vskip 1 truecm
\begin{tabular}{ccccc}
\multicolumn{1}{c}{Path} 
& \multicolumn{1}{c}{Integral}
& \multicolumn{1}{c}{Path}
& \multicolumn{1}{c}{Integral}
& \multicolumn{1}{c}{$\Phi$} \\ 
\hline
\hline
 $(0,0)\to (0,3)$ &  0.1011 & $(0,3) \to (2,3)$ & -0.3883 & -0.2872 \\
 $(0,0)\to (2,0)$ & -0.5274 & $(2,0) \to (2,3)$  & 0.2406 & -0.2868 \\
			&         &                   &         &         \\
 $(0,0)\to (0,0.8)$ &  0.1069 & $(0,0.8) \to (3,0.8)$ & -0.8036 & -0.6967 \\
 $(0,0)\to (3,0)$ & -1.4972 & $(3,0) \to (3,0.8)$ &  0.8 & -0.6972 \\
\end{tabular} 

\vskip 1 truecm

\caption[tabone]{
Entropy potential $\Phi$ computed by integrating along two different paths
in two different points of the $(\mu,\lambda)$ plane. The difference is
definitely smaller than the statistical error ($\approx 10^{-3}$).
}
\label{tab1}
\end{table}
\end{document}